\documentclass[journal=jpclcd,layout=twocolumn]{achemso}
\usepackage{graphicx}
\usepackage{amsmath}
\usepackage{amssymb}
\usepackage{dcolumn}
\usepackage{color}
\usepackage{natbib}
\usepackage{ctable}
\usepackage[sort&compress]{natbib}

\title{System size dependence of hydration shell occupancy}
\author{D. Asthagiri}
\email{Dilip.Asthagiri@rice.edu}
\affiliation{Department of Chemical and Biomolecular Engineering, Rice University, Houston, TX}
\author{Dheeraj Singh Tomar}
\email{dheerajstomar@gmail.com}
\affiliation{Akrevia Therapeutics, Cambridge, MA}

\begin{document}

\begin{abstract}
The free energies to evacuate the first hydration shell around a solute and a cavity defined by the first hydration shell depend on the system size. This observation interpreted within the quasichemical theory shows that both the hydrophilic and the hydrophobic contributions to hydration depend on the system size, decreasing with increasing system size. Although the net hydration free energy benefits somewhat from the balancing of hydrophilic and hydrophobic contributions, a large system still appears necessary to describe the effect of the solvent on the macromolecule. 
\begin{tocentry}
\includegraphics[scale=0.625]{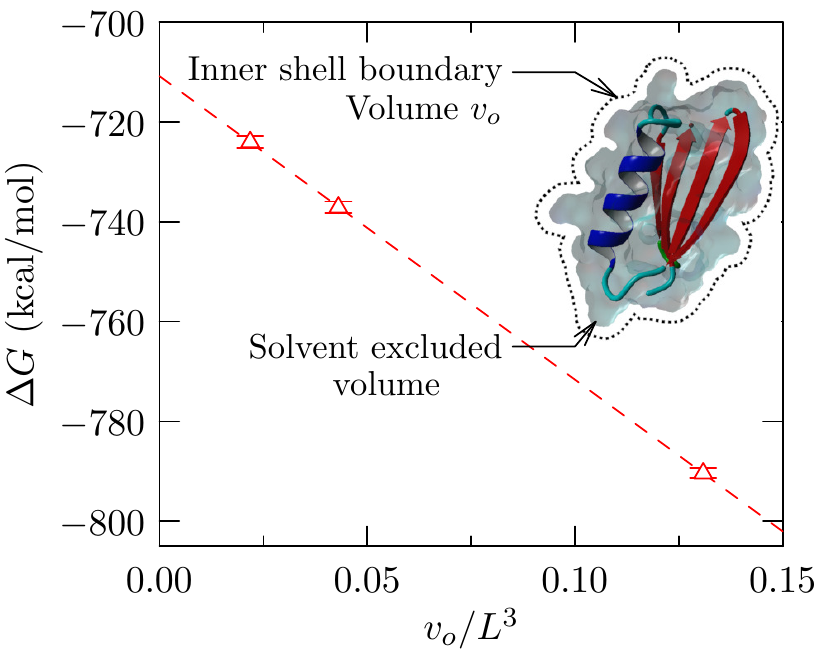} \\
Free energy to evacuate the inner-hydration shell of volume $v_o$ in a cubic
simulation cell of volume $L^3$. 
\end{tocentry}
\end{abstract}

How many water molecules are needed to simulate a hydrated protein? On the basis of their studies on hemoglobin, Karplus, Meuwly, and coworkers\cite{Meuwly:2018} have found that a large system size than conventionally used --- about 24 water molecules per protein heavy atom, or about 105,000 water molecules in total --- is required to capture the dynamics of deoxy-hemoglobin (PDB: 2DN2). While the statistical resolution of the results and the suitability of the analysis to address system size effects have  recently been questioned\cite{groot2019}, the observation in Ref.\ \citenum{Meuwly:2018} suggested to us a possible reason why our free energy calculations in conformational switching of protein $G_B$\cite{bryan:pnas09} proved inconclusive. In pursuing this line of research, we have uncovered a hitherto unexpected feature of finite size effect that we discuss below. 

In molecular simulations of hydration, using a finite number, $N$, of solvent molecules is unavoidable. In periodic simulations, there are \emph{implicit} system size effects due to periodicity, for example in pair-correlations \cite{haan:jcp81} and on electrostatic self-interaction \cite{Hummer:ions1996,Hummer:ions1998}. The are also \emph{explicit} finite size effects that arise from an ensemble dependence \cite{lebowitz:61,lebowitz:67}, for example, in calculating the compressibility \cite{egelstaff:96,velasco:99,trizac:2014}. From the vantage of quasichemical theory theory \cite{lrp:ES99,lrp:apc02,lrp:book,lrp:cpms}, we reasoned that such explicit finite size effects should be present in 
simulations of hydration as well. 

\begin{figure}[h]
\includegraphics[scale=0.6]{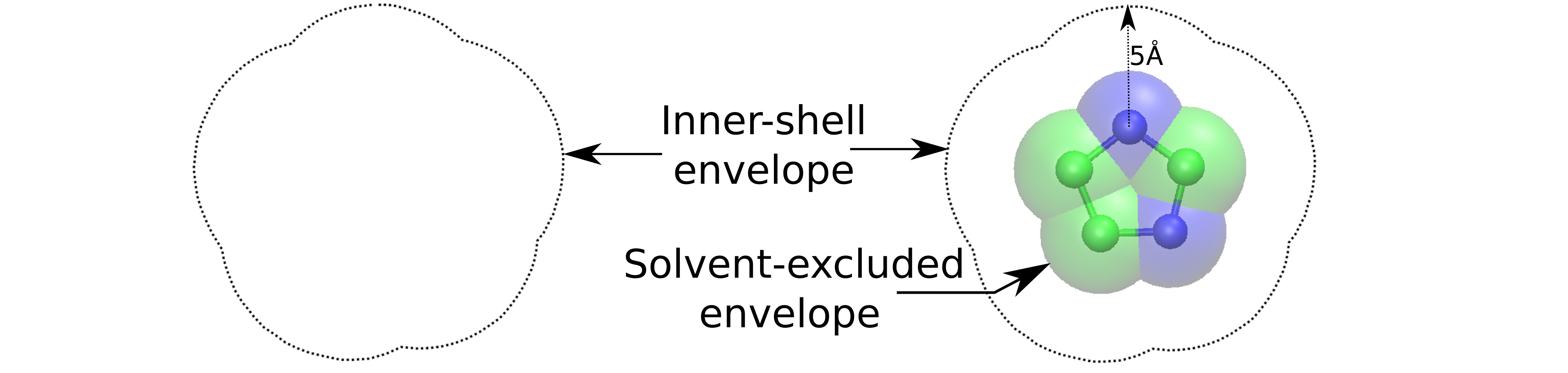}
\caption{The solute, an imidazole ring, is shown with its associated first hydration or inner shell, 
defined by the union of overlapping spheres of radius $\lambda = 5$~{\AA}, where $\lambda$ defines the distance between the heavy atom and water oxygen. (Hydrogens are not shown.) The inner-shell is \emph{open}, exchanging solvent with the bulk. Theory shows that $\lambda \approx 3$~{\AA} defines the solvent-excluded envelope.}\label{fg:imid}
\end{figure}

To anchor the discussion, first consider the hydration of a simple solute, an imidazole ring (Fig.~\ref{fg:imid}). 
Once we demarcate the inner-shell domain, the hydration free energy of the solute is given by quasichemical theory \cite{lrp:ES99,lrp:apc02,lrp:book,lrp:cpms,Weber:jctc12,tomar:bj2013,tomar:jpcb14,tomar:jpcb16,asthagiri:gly15,tomar:gdmjcp18,Tomar:2019a} as
\begin{eqnarray}
\mu^\mathrm{(ex)} = k_\mathrm{B}T \ln x_0  - k_\mathrm{B}T \ln p_0  + \mu^\mathrm{(ex)}( n=0) \, .
\label{eq:qct}
\end{eqnarray}
In Eq.~\ref{eq:qct}, $-k_\mathrm{B}T \ln x_0$ is the work done to evacuate the inner shell in the presence of the solute and $-k_\mathrm{B}T\ln p_0$ is the corresponding quantity in the absence of the solute; $x_0 (p_0)$ is the probability to observe an empty inner shell in the presence (absence) of the solute. $k_\mathrm{B} T\ln x_0$ is termed the \emph{chemistry} contribution, since it reflects the contribution to the hydration from short-ranged solute-solvent interactions. It is a measure of the hydrophilic contributions to hydration. $-k_\mathrm{B} T\ln p_0$ is the \emph{packing} contribution and is a measure of primitive hydrophobic effects\cite{Pratt:1992p3019,Pratt:2002p3001}.  $\mu^\mathrm{(ex)}(n=0) = k_\mathrm{B}T\langle e^{\beta\varepsilon}|n=0\rangle$ is the contribution to the hydration free energy from solute interaction with the solvent outside the inner-hydration shell; $\varepsilon$ is the solute-solvent interaction energy, and the averaging $\langle \ldots | n = 0\rangle$ is performed with solute-solvent thermally coupled, but with the inner-shell empty ($n=0$) of solvent molecules. As usual $\beta = 1/k_\mathrm{B}T$, where $T$ is the temperature and $k_\mathrm{B}$ is Boltzmann's constant. 

Our earlier studies\cite{tomar:bj2013,tomar:jpcb14,tomar:jpcb16,asthagiri:gly15,tomar:gdmjcp18,Tomar:2019a} establish $\lambda = 5$~{\AA} (Fig.~\ref{fg:imid}) is a conservative definition of the inner shell. For $\lambda \approx 3$~{\AA} the chemistry contribution is zero, uniquely identifying the domain 
excluded to the solvent. In our simulations, the inner-outer boundary is defined by a smooth, repulsive potential\cite{lrp:softcutoff}, but this choice is inconsequential for the discussion below. 

\begin{figure}[h!]
\begin{center}
\includegraphics[scale=0.975]{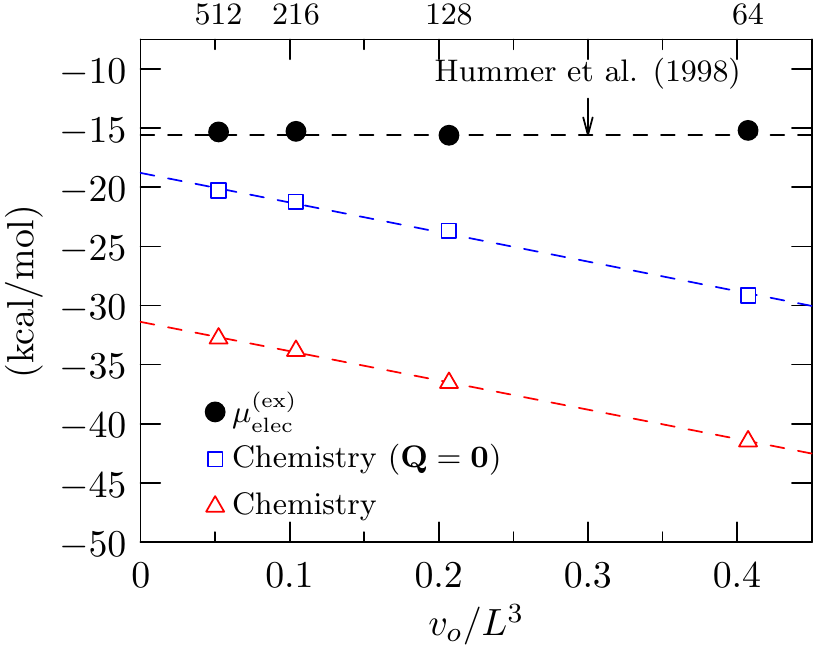}\hspace{5mm}\includegraphics[scale=0.95]{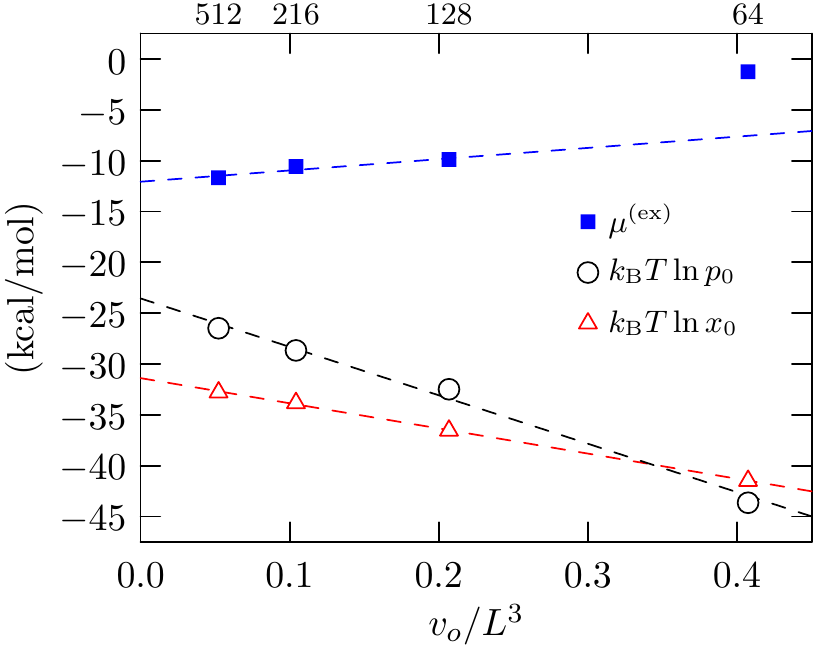}
 \end{center}
\caption{System size dependence of chemistry, packing, and hydration free energy of imidazole. Number of water molecules are noted atop each panel.
Statistical uncertainties ($1\sigma$) are about the size of the symbols. $v_o = 806.6$~{\AA}$^3$ is the volume of the inner-shell domain (supplemental information, SI).  $L^3$ is the volume of the cubic simulation system. The long-range contribution (SI) is not shown for clarity. \underline{Top}: $\mu^\mathrm{(ex)}_\mathrm{elec}$ from Eq.~\ref{eq:qct}.  The value reported by Hummer et al.\ (Fig.~5, Ref.\citenum{Hummer:ions1998}) is shown for comparison. Dashed curves are linear fits to the entire data set. \underline{Bottom}: System size dependence of the net hydration free energy. The extrapolated value is $-12.1$~kcal/mol. The dashed curves are linear fits excluding $N=64$. }
\label{fg:imidhyd}
\end{figure}

Previously, Hummer et al.\ \cite{Hummer:ions1998} had carefully examined electrostatic contributions in the hydration of imidazole and imidazolium. We follow the same simulation setup ($NVT$ ensemble) and same potential model, but obtain the charging free energy using Eq.~\ref{eq:qct}. The free energy of charging the imidazole is given
 by the difference of $\mu^\mathrm{(ex)}$ for imidazole and the analog with all charges set to zero ($\bf{Q=0}$). Since the packing contribution cancels in this difference, we need focus only on the chemistry term and the electrostatic contribution to the long-range term. Fig~\ref{fg:imidhyd} (top panel) shows that the chemistry contribution is 
 system size dependent, but the net electrostatic contribution is not. The excellent agreement with the value computed by Hummer et al.\ \cite{Hummer:ions1998} 
 confirms the internal consistency of our calculation. 

Fig~\ref{fg:imidhyd} (bottom panel) shows that the underlying chemistry (hydrophilic) and packing (hydrophobic) contributions are system size dependent, but these tend to compensate each other in the assessment of the net hydration free energy. However, the compensation is only partial. Thus, for example, for $N=216$, $\mu^\mathrm{(ex)} = -10.6\pm0.2$, whereas for $N=512$, $\mu^\mathrm{(ex)} = -11.7\pm0.2$.  Note also that for $N=64$ packing dominates chemistry, a behavior
found for hydrophobes\cite{asthagiri:jacs07,asthagiri:jcp2008}: this artifact arises due to a constant volume constraint (supplemental information, SI). As our results show, this unphysical behavior is masked in the calculation of $\mu^\mathrm{(ex)}_\mathrm{elec}$. 

The linearity of $\ln x_0$ and $\ln p_0$ with $v_o/L^3$ can be rationalized on the basis of two complementary perspectives. Analysis of fluctuations of number of solvent in a control volume (within a simulation cell) shows that a purely geometric factor of $(1-v_o/L^3)$ factors
from nonideal contributions \cite{velasco:99,trizac:2014}. This observation interpreted within a two-moment information theory model\cite{Hummer:1996p326,lrp:jpcb98} then shows that $\ln x_0$ or $\ln p_0$ should have a leading order dependence on  $v_ o/ L^3$. Alternatively, assuming ideal gas statistics gives the leading order dependence, we expect $\ln x_0$ (or $\ln p_0$) to depend as $\ln ( 1 - v_o / L^3) \approx -v_0/L^3$. 

We next consider the hydration of the 56-residue protein $G_B$ (PDB: 2LHD), which is one member of a pair of conformational switch peptides. The protein is also net neutral, helping minimize corrections that are necessary in treating the hydration of a charged molecule \cite{Hummer:ions1998}. ($G_B$ is remarkable, for a single mutation flips the conformation of the protein from the $4\beta+\alpha$ fold to a $3\alpha$ fold \cite{bryan:pnas09}.)

\begin{figure}[h!]
\begin{center}
\includegraphics[scale=1.0]{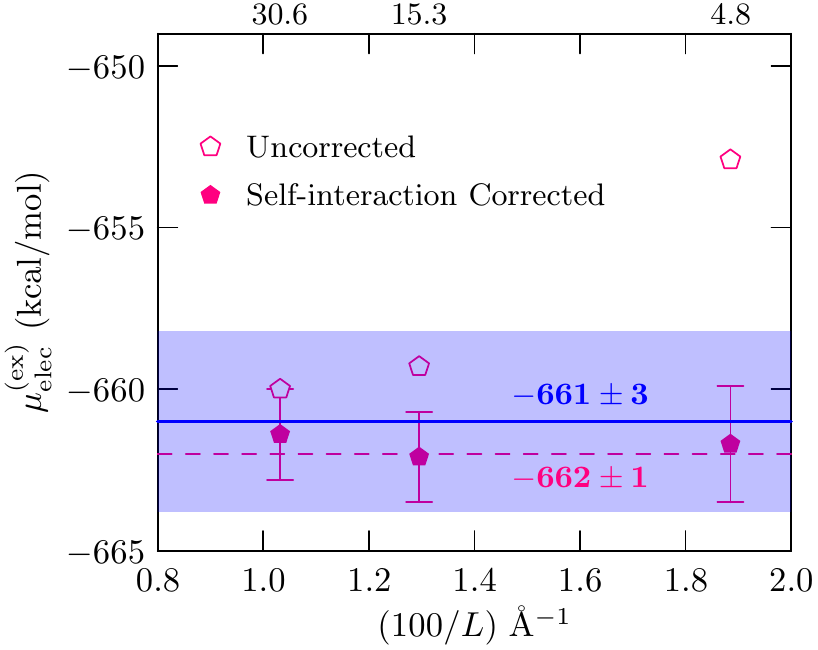}
 \end{center}
 \caption{The electrostatic contribution to the free energy of hydration of protein $G_B$. The open symbols are obtained without applying electrostatic self-interaction corrections, whereas the filled symbols include them (SI). The solid blue line is the result using Eq.~\ref{eq:qct}, with the blue shading indicating the uncertainty. Standard error of the mean is at $2\sigma$. Number of solvent molecules to the nearest thousand is indicated on top. For the largest system, there are about 68 solvent molecules per protein heavy atom.}\label{fg:gbelec}
 \end{figure}

We first establish the reference $\mu^\mathrm{(ex)}_\mathrm{elec}$ (Fig.~\ref{fg:gbelec}) for $G_B$ on the basis of standard coupling parameter integration (SI). 
Within statistical uncertainties, $\mu^\mathrm{(ex)}_\mathrm{elec}$ including appropriate corrections (SI) is independent of the system size. We do not include the electrostatic finite size correction because the leading order monopole contribution, zero for $G_B$, only has a $R^2 / L^3$ dependence \cite{Hummer:ions1998}, where $R$ is Born-radius of the solute. The results in Fig.~\ref{fg:gbelec} suggest that any finite size correction arising from the dipole contribution should be negligible. Further, just as we found for imidazole, $\mu^\mathrm{(ex)}_\mathrm{elec}$ obtained using the quasichemical approach is in excellent agreement with the  result based on coupling parameter integration. Having established this consistency, we next consider the individual packing and chemistry contributions. 
 
\begin{figure}[h!]
\begin{center}
\includegraphics[scale=0.95]{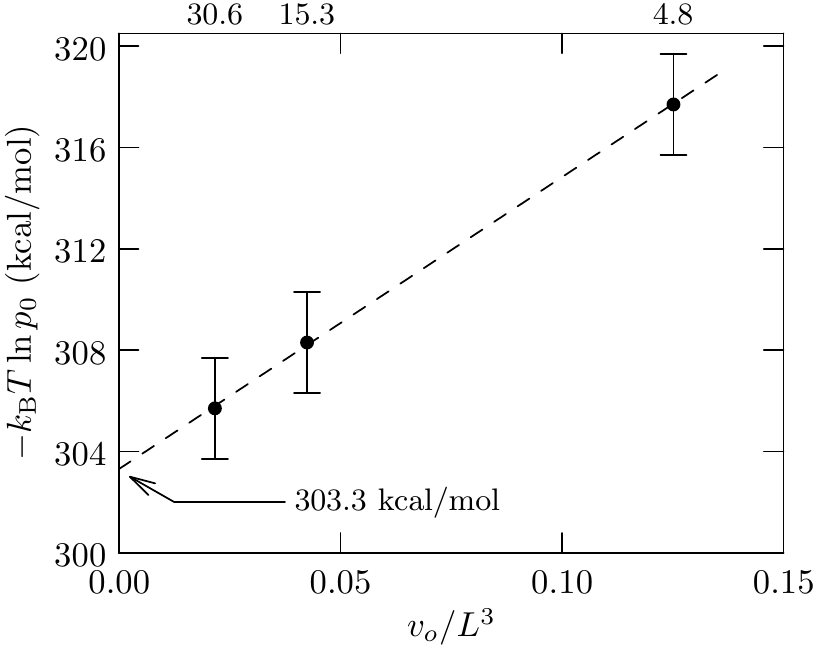}\hspace{5mm}\includegraphics[scale=0.95]{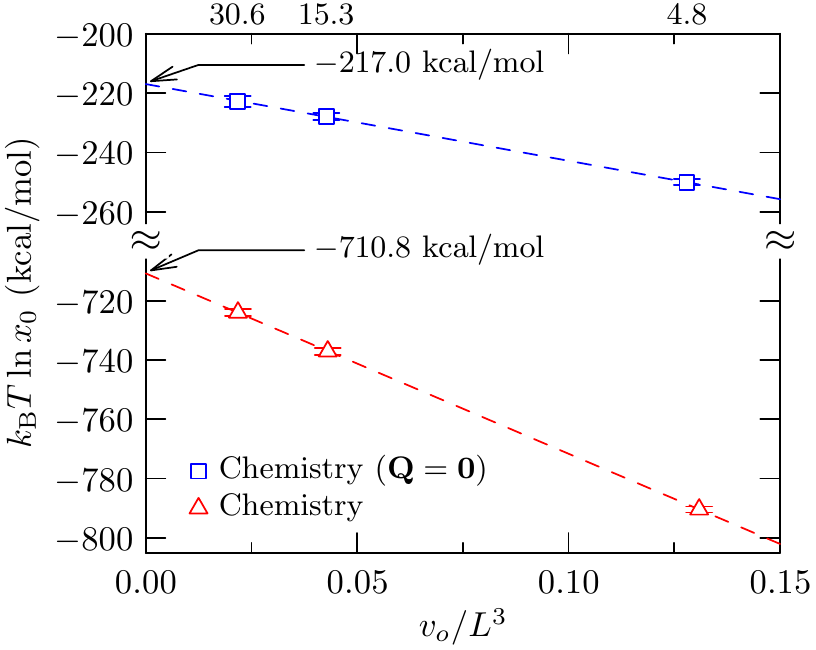}
 \end{center}
 \caption{\underline{Top}: The packing (primitive hydrophobic) contribution to $\mu^\mathrm{(ex)}$. \underline{Bottom}: 
 The chemistry contribution to $\mu^\mathrm{(ex)}$ for $G_B$ and its $\bf{Q=0}$ analog. $v_o = 20040.4$~{\AA}$^3$. Standard error of the mean is shown at $2\sigma$.}\label{fg:gbpackchem}
 \end{figure}

 Figure~\ref{fg:gbpackchem} (top panel) shows that  the (unfavorable) packing or primitive hydrophobic contribution becomes \emph{weaker} with increasing system size as does the (favorable) chemistry contribution (Fig.~\ref{fg:gbpackchem}, bottom panel).  These trends are similar to that for imidazole. 
 
\begin{figure}[h!]
\begin{center}
\includegraphics[scale=0.9]{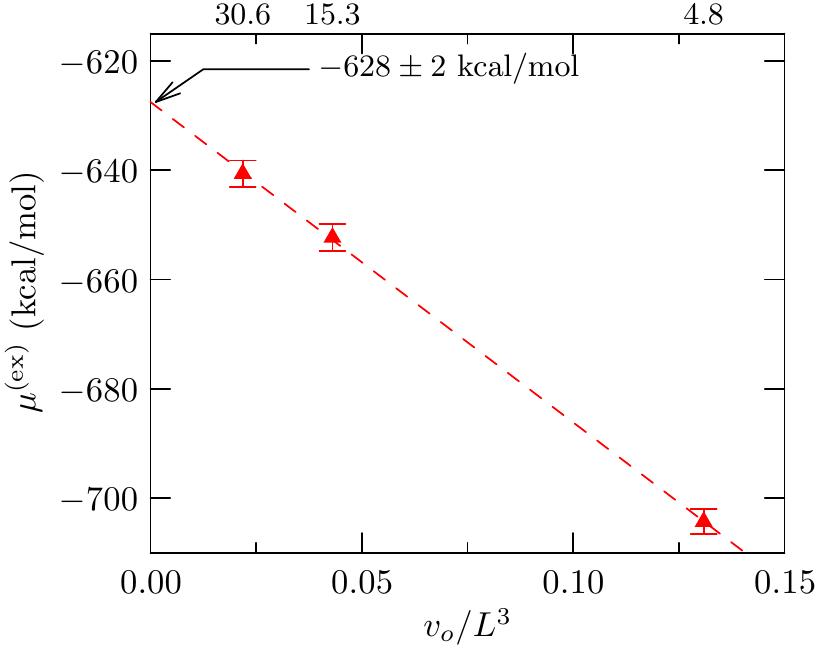}\hspace{5mm}\includegraphics[scale=0.9]{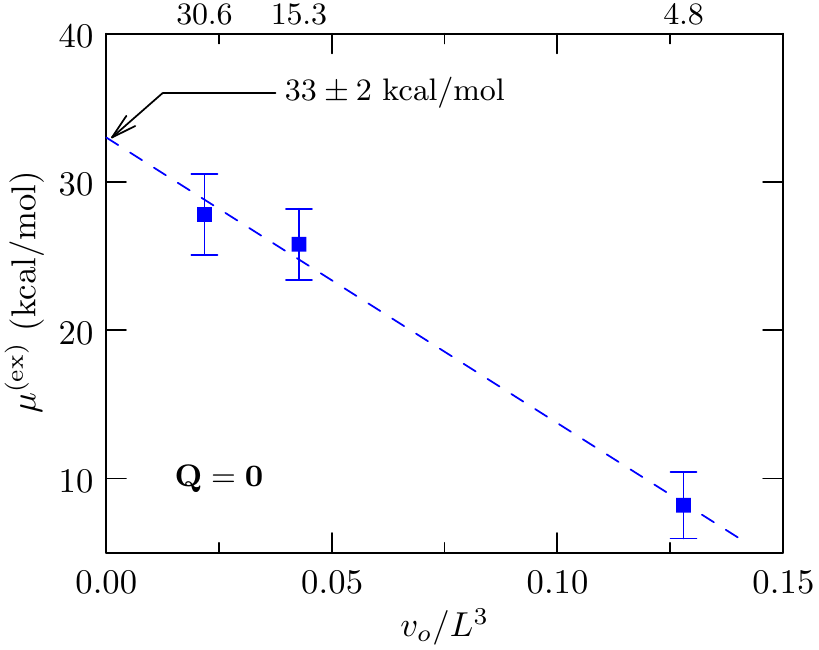}
 \end{center}
 \caption{Hydration free energy of $G_B$ (top) and its $\bf{Q=0}$ analog (bottom).  Electrostatic self-interaction corrections are 
 included in the free energy calculation of $G_B$.  Standard error of the mean is shown at $2\sigma$.}\label{fg:muex}
 \end{figure}

 Figure~\ref{fg:muex} gives the hydration free energy of the $G_B$ and its $\bf{Q=0}$ analog. The hydration free energy includes the long-range contribution, $\mu^\mathrm{(ex)}(n=0)$, which are only weakly dependent on the system size (SI). 
 From the extrapolated $L\rightarrow \infty$ value, we estimate the charging free energy $-628 - 33 = -661\pm 3\, (2\sigma)$~kcal/mol  noted in Fig.~\ref{fg:gbelec}.
 
The results above definitively establish that the first hydration shell occupancy, and hence the hydration thermodynamics itself, is sensitive to the size of the simulation system. The system size dependence noted here is quite general and arises because the first shell is an open system  that exchanges solvent with the bath. This feature, one that is made explicit by quasichemical theory \cite{lrp:ES99,lrp:apc02,lrp:book,lrp:cpms}, has implications for \emph{all simulation studies of solvation, in general, and hydration, in particular}. 

The present work suggests that the observations by Karplus, Meuwly, and coworkers\cite{Meuwly:2018} merits further study, but their rationalization 
of enhanced hydrophobic hydration in larger systems should be reconsidered. We find that the hydrophobic contribution is \emph{weakened} with increasing system size, 
consistent with the intuition of enhanced solvent density fluctuations at the scale of the observation volume in a larger system. However, the hydrophilic (chemistry) contribution becomes less favorable as well. That is the solvent is less effective in prying apart or loosening the protein structure, and this effect dominates the weakening of hydrophobic hydration, as our recent studies emphasize \cite{tomar:jpcb16,asthagiri:gly15,tomar:gdmjcp18,Tomar:2019a}.  This combined effect, which can be misinterpreted as a consequence of hydrophobic hydration\cite{Tomar:2019a}, should stabilize the conformation (assuming no change in intra-molecular interactions). 

The criticisms by Gapsys and de Groot also merit serious consideration. They emphasize that the analysis in Ref.~\citenum{Meuwly:2018} cannot satisfactorily address the issue of system size effect, but they do not foreclose the possibility of system size effects in kinetics and thermodynamics of conformational transitions in that study.  Importantly, for several model systems they find that the system size effect on the \emph{change} in free energy along a conformational transition is small. We suspect that \emph{relative} free energies of conformational transitions mask compensating effects of system size on hydrophobic and hydrophilic interactions noted here. Exploring this issue in biomolecular folding and assembly is left for future studies. 


\begin{acknowledgement}
We gratefully acknowledge computing support from National Energy Research Scientific Computing Center, which is supported by the Office of Science of the U.S. Department of Energy under Contract \# DE-AC02-05CH11231. D. A. thanks Lawrence Pratt and Walter Chapman for helpful comments and encouragement. We thank Vyatautas Gapsys and Bert de Groot for helpful comments and clarifications.
\end{acknowledgement}

\begin{suppinfo}
Information on the simulation systems, quasichemical calculations, calculation of reference value of charging free energy, tabulated data, and 
analog of Fig.~2 for $G_B$. 
\end{suppinfo}

 \bibliography{aife}

 \end{document}


\renewcommand{\thepage}{S\arabic{page}}  

\setcounter{section}{0}
\makeatletter
\renewcommand{\thesection}{S.\@Roman\c@section}

\setcounter{figure}{0}
\makeatletter
\renewcommand{\thefigure}{S\@arabic\c@figure}

\setcounter{table}{0}
\makeatletter
\renewcommand{\thetable}{S.\@Roman\c@table}

\setcounter{equation}{0}
\makeatletter
\renewcommand{\theequation}{S.\@arabic\c@equation}

\section{System}

All the simulations are performed using the NAMD code \cite{namd}. Since solute conformation was frozen and a rigid water model is used, we use a 2~fs time step for integrating the equations of motion.  

\subsubsection{Imidazole} 
The parameters for imidazole were obtained from Ref.\ \citenum{Hummer:ions1998}; consistent with that work, water was modeled using the SPC/E potential \cite{spce}. Following Ref.\  \citenum{Hummer:ions1998}, both neat water and water with one imidazole was modeled under NVT conditions. Neat water was simulated at a density of 0.997 gm/cc and the partial molar volume of imidazole was chosen as 2 times that of bulk water. The temperature of 298~K was enforced using a Langevin thermostat. The solvent box contained $N=64,128,256,\,\mathrm{or}\, 512$ molecules. 

For $N=64$, the Lennard-Jones interactions were switched to zero between 4.215~{\AA} and 5.215~{\AA}. For the other systems, the Lennard-Jones interactions were switched to zero between 6.63~{\AA} and 7.03~{\AA}. Electrostatic interactions were treated using particle mesh Ewald with a grid spacing of 0.5~{\AA}. 

\subsubsection{Protein $G_B$} 
The protein (PDB: 2LHD) was modeled using version 36m of the CHARMM forcefield \cite{charmm,c36one,c36two,c36m} and water was modeled using CHARMM-modified TIP3P potential \cite{tip32,tip3mod}. Charged residues were modeled in their standard ionization state at pH 7. The terminal residues were modeled in their ionized states. The protein is net neutral under the simulated conditions. 

First all the NMR structures in the PDB file were energy minimized: 500 steps fixing protein heavy atoms and a subsequent 500 steps without any constraints. 
We then calculate their combined intra-molecular plus solvation free energy (obtained using the GB/SA model \cite{still:jacs90,bashford:gb}), i.e.\ 
the free energy in solution. On this basis, we chose model 3, as it had the lowest free energy. 

The solvent box comprises $N=4775, 15302,\, \textrm{or}\, 30616$ waters. For all the systems, the Lennard-Jones interactions were switched to zero between 12~{\AA} and 13~{\AA} and electrostatic interactions were treated using particle mesh Ewald with a grid spacing of 0.5~{\AA}. All systems were modeled under NpT conditions, with the temperature of 298~K controlled using a Langevin thermostat and the pressure of 1~atm.\ controlled using a Langevin barostat \cite{feller:jcp95}. 

\section{Quasichemical theory}

The calculation of the hydration free energy components above closely follows earlier studies \cite{Weber:jctc12,tomar:bj2013,tomar:jpcb14,tomar:jpcb16,asthagiri:gly15,tomar:gdmjcp18,Tomar:2019a}. Here we present only the points needed to follow the main article. 
Once we demarcate the inner-shell region of size $\lambda$, $\mu^\mathrm{(ex)}$ is given by 
\begin{eqnarray}
\mu^{\mathrm (ex)} = k_\mathrm{B}T \ln x_0 (\lambda) - k_\mathrm{B}T \ln p_0 (\lambda) + \mu^{\mathrm (ex)} (n=0 | \lambda) \, .
\label{eq:qc}
\end{eqnarray}
We apply atom-centered fields, $\phi(r; \lambda)$, to carve a molecular cavity in the liquid or around the solute \cite{tomar:bj2013,tomar:jpcb14,tomar:jpcb16,asthagiri:gly15,tomar:gdmjcp18,Tomar:2019a}; $r$ is the distance to the oxygen atom of water and $\lambda$ is a parameter that defines the range of the field. We find that $\lambda \approx 5$~{\AA}  ensures that the conditional (i.e.\ $n=0 | \lambda$) binding energy distribution is gaussian to a good approximation. We denote this range as $\lambda_G$.  The largest value of $\lambda$, labelled $\lambda_{\rm SE}$, for which the chemistry contribution is zero has a special meaning. It demarcates the domain which is excluded to the solvent.  \textit{For the given forcefield and solute geometry, this surface is uniquely defined\cite{tomar:jpcb16}.} We find that $\lambda_{\rm SE} \approx 3$~{\AA}. With this choice, Eq.~\ref{eq:qc} can be rearranged as, 
\begin{eqnarray}
\mu^{\mathrm (ex)}  = \underbrace{ k_{\mathrm{B}} T \ln \left[x_0(\lambda_G)\frac{p_0(\lambda_{\rm SE})}{p_0(\lambda_G)}\right]}_{\rm revised\, chemistry}  \underbrace{-k_\mathrm{B} T \ln p_0(\lambda_{\rm SE})}_{\rm SE\, packing} +  \underbrace{\mu^{\mathrm(ex)} (n=0 | \lambda_G)}_{\rm long-range} \, .
\label{eq:qc1}
\end{eqnarray}
The term identified as revised chemistry has the following physical meaning. It is the free energy gained in allowing the solvent into the inner shell relative to the 
value for a solute that simply excludes the solvent.  This term explicates the role of short-range solute-solvent attractive interactions on hydration. Interestingly, the range between $\lambda_{SE} = 3$~{\AA} and $\lambda_G = 5$~{\AA} corresponds to the first hydration shell for a methane carbon\cite{asthagiri:jcp2008} and is an approximate descriptor of the first hydration shell of groups containing nitrogen and oxygen heavy atoms.  

For simplicity, in the main text we present results exclusively based on $\lambda_G$.  For the protein, the dependence of chemistry and revised chemistry on system size  is the same, as should be expected based on their physical meaning. However for imidazole, this is not be the case, since the calculations are under $NVT$ conditions. Specifically,
moving the solvent interface away from the solute costs substantially more \cite{merchant:jcp11a} under $NVT$ conditions than under $NpT$ conditions, 
skewing the relative balance of chemistry and packing. This is the reason why for $N=64$, packing dominates chemistry (Fig.~2, main text), a behavior that is unphysical for a polar compound. For the same reason, for imidazole simulated under $NVT$ conditions, chemistry and revised chemistry contributions have different dependencies on system size. 

\subsection{Chemistry and packing contributions}
To build the field to its eventual range of $\lambda_G = 5$~{\AA}, we progressively apply the field and for every unit {\AA} increment in the range, we compute the work using a seven-point Gauss-Legendre quadrature \cite{Hummer:jcp96}. Error analysis and error propagation was performed as before \cite{Weber:jctc12,tomar:bj2013,tomar:jpcb14,tomar:jpcb16,asthagiri:gly15,tomar:gdmjcp18,Tomar:2019a}: the standard error of the mean force was obtained using the Friedberg-Cameron algorithm \cite{friedberg:1970,allen:error} and in adding multiple quantities, the errors were propagated using  variance-addition rules. 

The starting configuration for each $\lambda$ point is obtained from the ending configuration of the previous point in the chain of states. For the packing contributions, a total of 35 Gauss points span $\lambda \in [0,5]$.  For the chemistry contribution, since solvent never enters $\lambda < 2.5$~{\AA}, we simulate $\lambda \in [2,5]$ for a total of 21 Gauss points. 

\subsubsection{Imidazole}
We perform 1~ns of simulation at each $\lambda$ and use the data from the last 0.5~ns for analysis. Force data is archived every 50~fs for analysis. 

\subsubsection{$\bf{G_B}$}
We perform 0.8~ns of simulation at each $\lambda$ and use the data from the last 0.4~ns for analysis. Force data is archived every 50~fs for analysis. 

\begin{figure}
\includegraphics[scale=0.5]{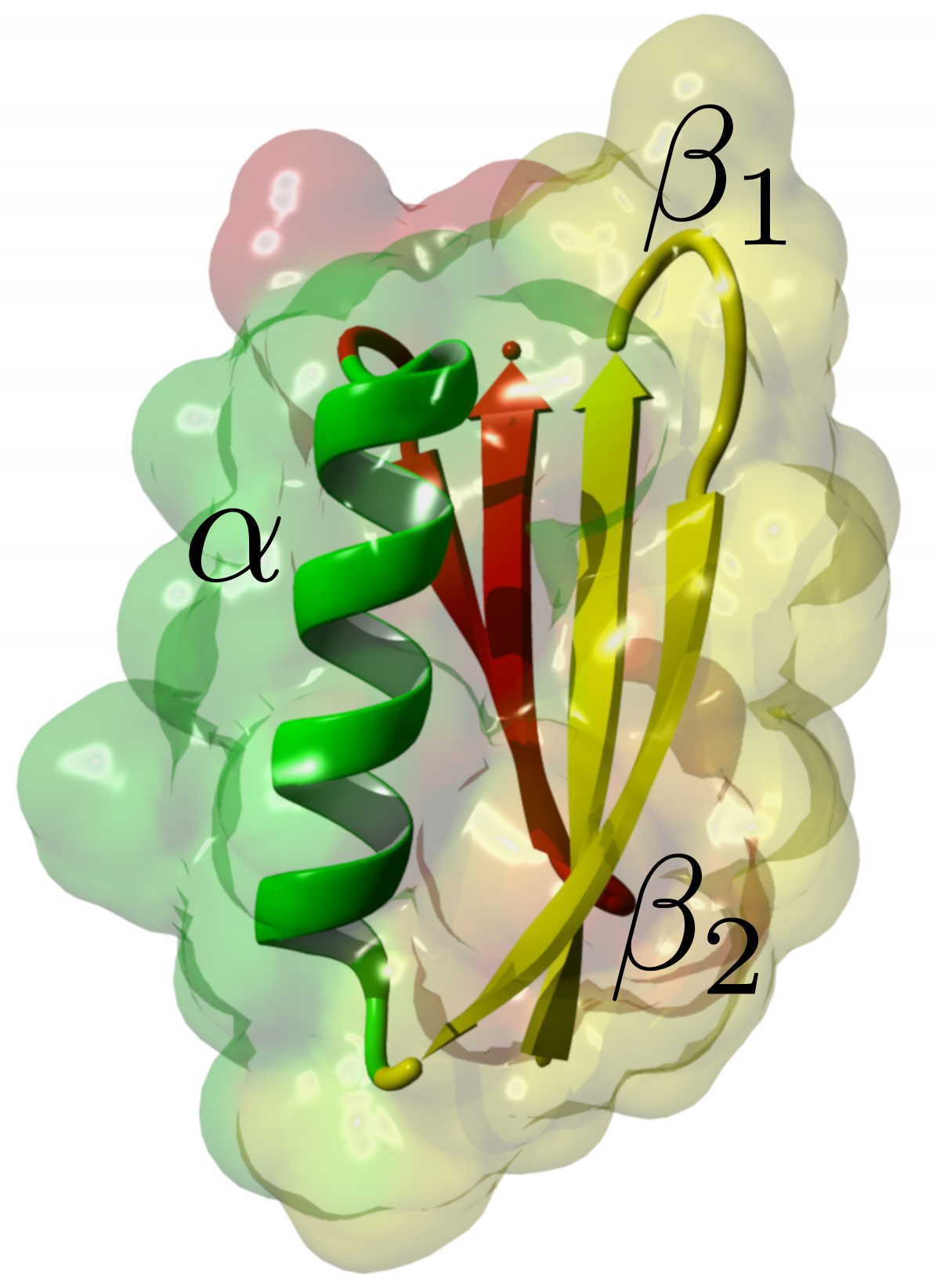}
\caption{The structure of protein $G_B$ showing the main secondary structural elements.}\label{fg:gb}
\end{figure}

The packing calculation for $G_B$ includes additional subtleties. To prevent water molecules from being trapped inside the inner-shell envelope, we do the packing calculation in three stages, corresponding to the $\beta_1$, $\alpha$, and $\beta_2$ secondary structural domains of the protein (Fig.~\ref{fg:gb}). 
We first create the $\lambda_{SE}$ cavity for $\beta_1$. Next, with the $\beta_1$ cavity in place, we create the $\lambda_{SE}$ cavity for 
$\alpha$. Finally, we create the $\beta_2$ cavity in the presence of the $\lambda_{SE}$ cavities for $\beta_1$ and $\alpha$.  With the 
$\lambda_{SE}$ cavity for the entire protein in place, we create the $\lambda_G$ cavity.  Such an approach also provides information on the 
energetics of conditional cavity formation to accommodate each secondary structural element and proves helpful in monitoring convergence of the calculations. 

\subsection{Long-range contribution}

Throughout, solute-solvent binding energies were obtained using the {\sc PairInteraction} module in NAMD.

The conditional solute-solvent binding energy distribution is $P(\varepsilon| n = 0)$, where $\varepsilon$ is the solute-solvent binding energy. For $P(\varepsilon|n=0)$ described by a Gaussian, we have  \cite{lrp:book,lrp:cpms}
\begin{eqnarray}
\mu^\mathrm{ex}(n=0) & = & \langle \varepsilon | n=0 \rangle + \frac{\beta}{2}\sigma^2 
\label{eq:gaussian}
\end{eqnarray}
In the above equations, $\langle \varepsilon | n=0 \rangle$ is the mean binding energy
and $\sigma^2$ is the variance of the distribution, with solvent prohibited from entering the inner shell. For characterizing  $P(\varepsilon| n=0)$, the starting configuration for the $\lambda_G = 5$~{\AA} simulation was obtained from the endpoint of the Gauss-Legendre procedure for the chemistry calculation.

\subsubsection{Imidazole}
 The system with $\lambda_G$ was simulated for 2~ns and frames were saved every 200~fs. We used the last 9500 frames for analysis. The long-range contribution includes both electrostatics and van~der~Waals interactions. The individual distributions are also Gaussian, serving as a further consistency check of the simulations (see, for example, Ref.~\citenum{tomar:bj2013}). 

\subsubsection{$\bf{G_B}$}
 Since the protein has a very large dipole moment ($\approx 150$~D), we use the Gaussian model only for van~der~Waals interactions. 
 The $\bf{Q=0}$ system with $\lambda_G$ was simulated for 2~ns and frames saved every 200~fs. We used the last 9000 frames for calculating the van~der~Waals contribution
using the Gaussian model.  For the electrostatic contributions, we use a 3-point quadrature rule \cite{Hummer:jcp96}. At each Gauss point, the system was simulated for 0.7~ns and frames saved every 200~fs for further analysis.  In contrast to our earlier studies on simpler peptides \cite{tomar:jpcb16,asthagiri:gly15,tomar:gdmjcp18,Tomar:2019a}, we find that the linear response result (the Gaussian model result) differs from the 3-point result by between $3$ and $4$~kcal/mol, outside the uncertainty of $0.2$~kcal/mol  for the quadrature-based result (see tables below).

\section{Reference free energy of charging}

For charging a solute from $\bf{Q=0}$ to the final charge distribution $\bf{Q}$, we use well-documented ideas\cite{Hummer:ions1998,Hummer:jcp96}. 
Let $\phi(\bf{r}_{\alpha\beta})$ be the Ewald potential between partial charges $q_\alpha$ and $q_\beta$ at locations $\bf{r}_\alpha$ and $\bf{r}_\beta$ in the solute. (Please note that the solute is treated as a rigid entity in our simulations.) Then the free energy of charging the solute is given by\cite{Hummer:ions1998} 
\begin{eqnarray}
\mu^{\mathrm{(ex)}} = \mu^{\mathrm{(ex)}}_\mathrm{sim} + \frac{1}{2} \sum\limits_{\alpha,\beta \atop \alpha\neq\beta} q_\alpha q_\beta \left[ \phi(\bf{r}) - \frac{1}{|\bf{r}_{\alpha\beta}|}\right] + \frac{1}{2}\sum_\alpha q_\alpha^2 \xi \, 
\label{eq:ewald}
\end{eqnarray}
where $\xi = -2.827297 / L$ is the Wigner potential for a cubic cell of length $L$. The second and third terms together constitute the self-interaction correction noted in the main text. The second term accounts for the interaction between partial charge sites on the solute and its periodic images. The third term is the ionic self-interaction contribution. $\mu^\mathrm{(ex)}_\mathrm{sim}$ is the contribution to the hydration free energy from solute interaction with the solvent, and it is this term that is obtained using quadratures. For constant pressure simulations, the self-interaction contribution should be averaged as the simulation volume fluctuates. In practice, for the box sizes considered in this work, 
we find that using a box size based on the average volume suffices in computing the correction. (While strictly not correct, this procedure has the virtue of simplicity and 
the error due to this simplification is insignificant.) The self-interaction contributions were obtained using an in-house Ewald summation code; the Ewald screening parameter and number of k-space vectors follow the recommendation in Ref.\ \citenum{Hummer:ions1998}.

For imidazole, we took the reference free energy value from Ref.~\citenum{Hummer:ions1998}. For $G_B$, at each Gauss point we simulated the system for 0.6~ns, saving data every 200~fs. We used the last 2500 frames for analysis. 

\section{Molecular volumes}

To compute the volume associated with an envelope defined by $\lambda$, we assign the radius $\lambda$ to the solute heavy atoms. We then use the MSMS code \cite{msms} to calculate the \emph{solvent excluded} volume for that envelope.

\newpage
\section{Data}

Throughout, energy values are reported in kcal/mol. 

\renewcommand\arraystretch{1.5}

\subsection{Imidazole}

\begin{table}[h]
\caption{Packing contribution for $\lambda_G = 5$~{\AA} and $\lambda_{SE} = 3$~{\AA}. Standard error of the mean is at $1\sigma$.}
\begin{tabular}{crrr}
\toprule
\multicolumn{2}{c}{} & \multicolumn{2}{c}{Packing} \\ \cline{3-4}
N & $L$ ({\AA}) & \multicolumn{1}{c}{$\lambda_G$} & \multicolumn{1}{c}{$\lambda_{SE}$} \\ \midrule
512	& 24.859	&  $26.5\pm 0.1$ & $6.03\pm 0.05$ \\
256	& 19.730	& $28.7\pm 0.2$ & $6.09\pm	0.05$ \\
128	& 15.660	& $32.5\pm 0.1$ & $6.14\pm	0.07$ \\
64	& 12.429	& $43.6\pm 0.2$ & $6.45\pm	0.05$ \\
\bottomrule
\end{tabular}
\end{table}

\begin{table}[h]
\caption{Chemistry and long-range contributions for the $\bf{Q}$ and $\bf{Q=0}$ cases. Elec$_{corr}$ is the Ewald self-interaction correction. Standard error of the mean is at $1\sigma$.}
\begin{tabular}{crrrrrr}
\toprule
 \multicolumn{1}{c}{} & \multicolumn{1}{c}{} & \multicolumn{2}{c}{Chemistry} & \multicolumn{1}{c}{} & \multicolumn{2}{c}{Long-range} \\ \cline{3-4}\cline{6-7}
N & $L$ ({\AA}) & \multicolumn{1}{c}{$\bf{Q}$} & \multicolumn{1}{c}{$\bf{Q=0}$} & Elec$_{corr}$ &  \multicolumn{1}{c}{$\bf{Q}$}  & \multicolumn{1}{c}{$\bf{Q=0}$}  \\ \midrule 
512 &	24.891 &	$-32.8\pm 0.1$ &	$-20.3\pm 0.1$ & $-0.06$ & $-5.27\pm 0.06$  & $-2.56$ \\
256 &	19.782 &	$-33.8\pm 0.1$ &	$-21.2\pm 0.1$ & $-0.12$ & $-5.25\pm 0.06$  & $-2.73$ \\
128 &	15.741 &	 $-36.5\pm 0.1$ &	$-23.7\pm 0.1$ & $-0.24$ & $-5.55\pm 0.05$  & $-3.07$\\
64  & 	12.557 &	 $-41.5\pm 0.1$ & 	$-29.2\pm 0.1$ & $-0.48$ & $-2.92\pm 0.05$ &	$-0.55$\\
\bottomrule
\end{tabular}
\end{table}
%

\newpage
\subsection{$\bf{G_B}$}
\begin{table}[ht]
\caption{Free energy of charging $G_B$ from $\bf{Q=0}$ to the final partial charge distribution $\bf{Q}$ using a 3-point Gauss-Legendre quadrature. $\mu^\mathrm{(ex)}_\mathrm{sim}$ is the solute-solvent interaction contribution to the free energy. Elec$_{corr}$ is the self-interaction correction (Eq.~\ref{eq:ewald}). The given box length is the box length based on the average volume for the system at the final charge distribution. The correction is evaluated using this value of the box length. Standard error of the mean is at $1\sigma$.}\label{tb:gbcharging}
\begin{tabular}{crrrr}
\toprule
N & $L$ ({\AA}) & \multicolumn{1}{c}{$\mu^\mathrm{(ex)}_\mathrm{sim}$} & Elec$_{corr}$ & \multicolumn{1}{c}{$\mu^\mathrm{(ex)}_\mathrm{elec}$} \\ \midrule 
30616 & 96.9 & $-660.0 \pm 0.7$ & $-1.4$ & $-661.4\pm 0.7$ \\ 
15302 & 77.2 & $-659.3 \pm 0.7$ & $-2.8$ & $-662.1\pm 0.7$ \\
4775   & 53.1 & $-652.9 \pm 0.9$ & $-8.8$ & $-661.7\pm 0.9$ \\ 
\bottomrule
\end{tabular}
\end{table}

\begin{table}[h]
\caption{Packing contribution for $\lambda_G = 5$~{\AA} and $\lambda_{SE} = 3$~{\AA}. Standard error of the mean is at $1\sigma$.}\label{tb:gbPacking}
\begin{tabular}{crrr}
\toprule
\multicolumn{2}{c}{} & \multicolumn{2}{c}{Packing} \\ \cline{3-4}
N & $L$ ({\AA}) & \multicolumn{1}{c}{$\lambda_G$} & \multicolumn{1}{c}{$\lambda_{SE}$} \\ \midrule
30616 &	97.41 &	$305.7\pm 1.0$	 & $218.5\pm 1.0$ \\
15302 &	77.85 &	$308.3\pm 1.0$	 & $219.3\pm 1.0$ \\
4775	   &     54.3 &	$317.7\pm 1.0$ &  $222.2\pm 1.0$ \\
\bottomrule
\end{tabular}
\end{table}

 \newpage
\begin{table}[h]
\caption{Chemistry and long-range electrostatic (Elec.) and van~der~Waals (vdW) contributions in the net hydration free energy, $\mu^\mathrm{(ex)}$, of $G_B$. The long-range electrostatic contribution is obtained using a 3-point Gauss-Legendre quadrature. The linear response (or gaussian) value for the electrostatic contribution is noted in parenthesis. Standard error of the mean is at $1\sigma$.}\label{tb:gbChemistry}
\begin{tabular}{crrrrrr} 
\toprule
    &                    &                  &  \multicolumn{2}{c}{Long-range} &     &   \\ \cline{4-5} 
N & $L$ ({\AA}) & Chemistry &  \multicolumn{1}{c}{Elec.} & \multicolumn{1}{c}{vdW} & Elec$_{corr}$ & $\mu^\mathrm{(ex)}$ \\ \midrule
30616 &	97.2	& $-724.0\pm 0.6$ & $-165.8\pm 0.2$ ($-161.8$) & $-55.1\pm 0.2$ & $-1.4$	& $-640.6$ \\
15302 &	77.5	& $-737.1\pm 0.6$ &	 $-166.1\pm 0.3	$ ($-162.3$) & $-54.6	\pm 0.3$ & $-2.8$	& $-652.3$ \\
4775	  &	53.5	& $-790.4\pm 0.5$ &  $-163.4\pm 0.2$ ($-160.3$) &	$-59.4\pm	0.1$	& $-8.8$	& $-704.3$ \\
\bottomrule
\end{tabular}
\end{table}
Tables~\ref{tb:gbPacking} and~\ref{tb:gbChemistry} can be used to calculate the revised chemistry contribution. For example, for $N=4775$, the revised chemistry contribution 
is $-790.4 + 317.7 - 222.2 = -694.9$~kcal/mol. This is the work done in allowing water molecules to enter the inner-shell of $G_B$ relative to the case when the repulsive hard-core solute. For $G_B$, the revised chemistry contribution has the same qualitative dependence on $v_o / L^3$ as the chemistry contribution (Fig.~4, main text).

\begin{table}[h]
\caption{Chemistry and long-range (vdW) contributions in the net hydration free energy, $\mu^\mathrm{(ex)}$, of the $\bf{Q=0}$ analog of $G_B$. Standard error of the mean is at $1\sigma$.}\label{tb:gbQoChemistry}
\begin{tabular}{crrrr} 
\toprule
N & $L$ ({\AA}) & Chemistry & \multicolumn{1}{c}{vdW} & $\mu^\mathrm{(ex)}$ \\ \midrule
30616 &	97.3	& $-222.8\pm 0.9$  & $-55.1\pm 0.2$ 	& $27.8$ \\
15302 &	77.7	& $-227.9\pm 0.6$  & $-54.6 \pm 0.3$ 	& $25.8$ \\
4775	  &	53.9	& $-250.1\pm 0.5$  & $-59.4\pm 0.1$	& $8.2$ \\
\bottomrule
\end{tabular}
\end{table}

\newpage
\subsection{Free energy of charging $G_B$}

In the main text, we obtain the free energy of charging $G_B$ from the difference of the (absolute) hydration free energies of $G_B$ and its $\bf{Q=0}$ analog. The absolute
hydration free energies were obtained from the $L\rightarrow\infty$ extrapolation, and using those values we had the result $\mu^\mathrm{(ex)}_\mathrm{elec} = -628 - 33 = -661\pm 3\, (2\sigma)$~kcal/mol. 

\begin{figure}[h!]
\includegraphics{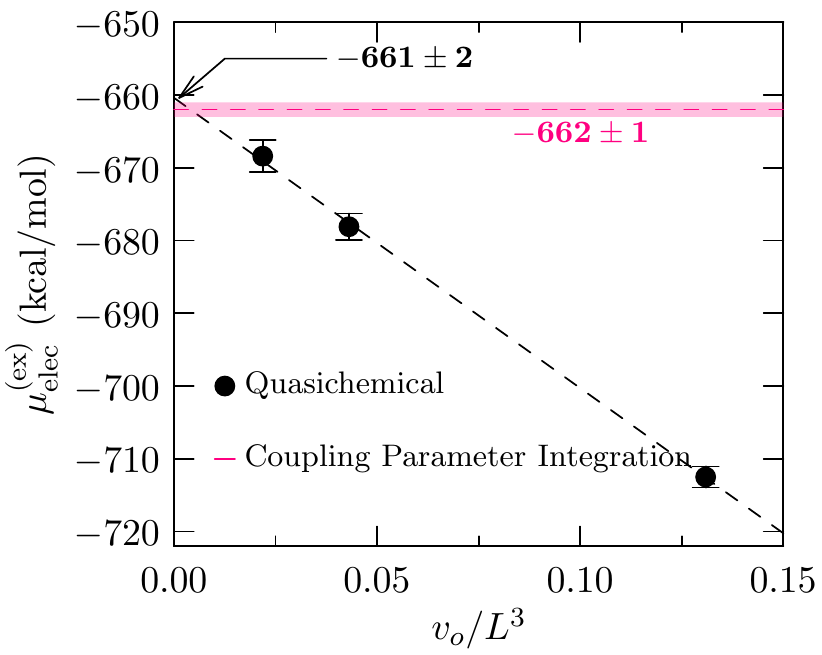}
\caption{System size dependence of the charging free energy for $G_B$. The quasichemical value includes self-interaction corrections.}\label{fg:dchem}
\end{figure}

As we did for imidazole, we can also find the charging free energy from the difference of the chemistry contributions plus the long-range electrostatic and self-interaction correction values (Table~\ref{tb:gbChemistry}). This calculation also proves illuminating. As Fig.~\ref{fg:dchem} shows, the $L\rightarrow \infty$ value is in excellent agreement with the average value obtained from coupling parameter integration procedure (Table~\ref{tb:gbcharging}). 

But there is an important contrast to me made with imidazole. Unlike the case for imidazole (Fig.~2), even after including self-interaction corrections, we find that the 
charging free energy (from the quasichemical procedure) has a strong system size dependence. There are two reasons for this: (1) the calculations are under $NpT$ conditions and (2) in the charging of the solute with the empty inner-shell, we will likely need to consider electrostatic \emph{finite size} corrections, since the solute with an empty $\lambda = 5$~{\AA} envelope will occupy a much larger volume in the simulation cell. (The volume of the $\lambda_G$ envelope is nearly 1.8 times 
that of the $\lambda_{SE}$ envelope.)  For an ion of Born radius $R$, Hummer et al.\ (Ref.~\citenum{Hummer:ions1998}) showed that the \emph{finite size} correction will scale as $R^2 / L^3$. Analytically working out the scaling for a protein with its partial charge distribution is an daunting task, but it may be possible to estimate the correction numerically (by solving Poisson's equation). While we have not attempted this, by analogy with the ion case we expect the correction to be a positive contribution that is inversely proportional to the box volume, i.e. the quasichemical value should be below the coupling parameter value. The trend in Fig.~\ref{fg:dchem} is consistent with this expectation. Further exploration of the finite size correction is left for future studies.

\newpage 

\providecommand{\latin}[1]{#1}
\makeatletter
\providecommand{\doi}
  {\begingroup\let\do\@makeother\dospecials
  \catcode`\{=1 \catcode`\}=2 \doi@aux}
\providecommand{\doi@aux}[1]{\endgroup\texttt{#1}}
\makeatother
\providecommand*\mcitethebibliography{\thebibliography}
\csname @ifundefined\endcsname{endmcitethebibliography}
  {\let\endmcitethebibliography\endthebibliography}{}